\documentclass{elsart}

\usepackage{graphics}
\usepackage{epsfig}

\usepackage{amssymb}

\begin{document}
\begin{frontmatter}



\title{Extrinsic inhomogeneity effects in magnetic, transport
and magnetoresistive properties of La$_{1-x}$Ca$_{x}$MnO$_{3}$
($x\approx 0.33$) crystal prepared by the floating zone method}

\author[ILTPE]{B. I. Belevtsev\corauthref{cor}},
\author[TEXAS]{D. G. Naugle},
\author[TEXAS]{K. D. D. Rathnayaka},
\author[TEXAS]{A. Parasiris},
\author[IPPAN]{J. Fink-Finowicki}

\address[ILTPE]{B. Verkin Institute for Low Temperature Physics and
Engineering, National Academy of Sciences, Kharkov, 61103,
Ukraine}

\address[TEXAS]{Texas A\&M University, College Station, TX 77843, USA}

\address[IPPAN]{Institute of Physics, Polish Academy of Sciences,
32/46 Al. Lotnikov, 02-668 Warsaw, Poland}
\corauth[cor]{Corresponding author. Email address:
belevtsev@ilt.kharkov.ua. Fax: ++380-572-335593. Phone:
++380-572-308563.}

\begin{abstract}
The magnetic, transport and magnetoresistive properties of
La$_{1-x}$Ca$_{x}$MnO$_{3}$ ($x\approx 0.33$) crystals prepared by
the floating-zone method are studied. In general, these properties
testify to rather good crystal perfection of the sample studied.
In particular, a huge magnetoresistance ($ [R(0)-R(H)]/R(H)$ in
the field $H=$~5 T is about $\approx 2680$~\%) is found near the
Curie temperature (216 K). At the same time, some distinct
features of measured properties indicate the influence of
extrinsic inhomogeneities arising due to technological factors in
the sample preparation. Analysis of the data obtained shows that
these are rare grain boundaries and twins.
\end{abstract}

\begin{keyword}
manganites \sep  colossal magnetoresistance \sep magnetically
ordered materials \sep magnetic inhomogeneities

\PACS 75.47.Gk; 75.47.Lx

\end{keyword}
\end{frontmatter}

\section{Introduction}
\label{int}
 The structural, magnetic and electron transport
properties of mixed-valence manganites of the type
R$_{1-x}$A$_x$MnO$_3$ (where R is a rare-earth element, A a
divalent alkaline-earth element) attracted much attention in the
scientific community in the last decade (see reviews
\cite{ramirez,coey,dagotto,nagaev,kim,dagotto1}). The interest is
caused by observation of huge negative magnetoresistance (MR) near
the Curie temperature, $T_c$, of the paramagnetic-ferromagnetic
transition for manganites with $0.2 \leq x \leq 0.5$. This
phenomenon called ``colossal'' magnetoresistance (CMR), is
expected to have application in advanced technology. The unique
properties of mixed-valence manganites are determined by the
complex spin, charge and orbital ordered phases, and, therefore,
are of great fundamental interest for physics of strongly
correlated electrons. At present, it is believed that one of the
key features of manganites is their intrinsic inhomogeneities in
the form of coexisting, competing ferromagnetic and
antiferomagnetic/paramagnetic phases
\cite{dagotto,nagaev,dagotto1}. This phenomenon is generally
called ``phase separation''.  The known experimental studies give
numerous (but predominantly indirect) evidence of structural and
magnetic inhomogeneities in manganites, but it hardly can be said
that they are intrinsic in all cases. The point is that in all
manganites extrinsic inhomogeneities are inevitably present (even
in single crystal samples). The inhomogeneities of this type arise
due to various technological factors in the sample preparation.
They can cause chemical-composition inhomogeneity (first of all in
the oxygen content), structural inhomogeneities (polycrystalline
or even granular structure), strain inhomogeneities and so on. It
is easy to find in the literature many experimental studies in
which finding of phase separation effects is proclaimed, but the
interpretations are often doubtful. In such cases the effects of
technological inhomogeneities are quite obvious or, at least,
cannot be excluded.
\par
The experimental data demonstrate that technological
inhomogeneities are unavoidable for any preparation method
\cite{boris2}. For this reason, in many cases it is better to
speak about multiphase coexistence instead of the phase
separation.  It is quite reasonable, therefore, that consideration
of experimental data for mixed-valence manganites and development
of theoretical models for them take into account the unavoidable
influence of extrinsic disorder and inhomogeneity. These
inhomogeneities can act separately as well as together with the
intrinsic inhomogeneities (phase separation) and determine to a
great extent the magnetic and magnetotransport properties of these
compounds.
\par
In this article we present a study of structure, magnetic, and
magnetoresistive properties of crystals with nominal composition
La$_{0.67}$Ca$_{0.33}$MnO$_3$ prepared by the floating zone
method. The measured properties generally indicate that the sample
studied is of rather high crystal perfection. At the same time
some features of the transport properties reflect the essential
influence of extrinsic structural inhomogeneities arising due to
technological factors in the sample preparation.

\section{Experimental}
Crystals of nominal composition La$_{0.67}$Ca$_{0.33}$MnO$_3$ were
grown by the floating zone method at the Institute of Physics,
Warsaw. The appropriate amounts of starting materials La$_2$O$_3$,
CaCO$_3$ and MnO$_2$ were calcined at 1000$^{\circ}$C, then mixed,
compacted into pellets and sintered at 1400$^{\circ}$C.
Subsequently, the pellets were milled, and the resulting powder
was then pressed to form the feed rod with a diameter of 8 mm and
a length of 90 mm. This rod was further sintered at
1470$^{\circ}$C for 12 h in air. An optical furnace (type
URN-2-3Pm made by Moscow Power Engineering Institute) with two
ellipsoidal mirrors and a 2500 W xenon lamp as the heat source was
used for crystallization. The feed rod and the growing crystal
were rotated in opposite directions to make heating uniform and to
force convection in the melting zone. The growth rate was 1 mm/h.
An additional afterheater was used for lowering the temperature
gradients in the growing crystal.
\par
As shown by previous experience \cite{fink}, manganite crystals
produced by this technique are almost single crystals, and, in
this respect, they have far better crystal quality and far less
porosity than samples prepared by a solid-state reaction
technique. On the other hand, the crystals have twin domain
structure and may have a few grain boundaries. This question will
be discussed more thoroughly below.
\par
The disc-shaped sample for investigation was cut from the middle
part of the cylindrical crystal (about 8 mm in diameter). It is
known \cite{muk1,muk2} that La$_{1-x}$Ca$_x$MnO$_3$ crystals grown
by the floating zone technique have an inhomogeneous distribution
of La and Ca along the growth direction. The initial part of the
crystal is usually somewhat enriched in La and depleted in Ca. The
microprobe elemental analysis has shown that the sample studied
has a chemical composition close to the nominal one. This is
supported by its transport and magnetic properties as well, as
described below. The sample characterization and measurements were
done by a variety of experimental techniques. Crystal structure
was determined by the X-ray diffraction (XRD) method. The XRD
spectra (from powders and bulk pieces taken from the crystal) were
obtained using a Rigaku model D-MAX-B diffractometer with a
graphite monochromator and Cu $K_{\alpha 1,2}$ radiation. The ac
susceptibility and dc magnetization were measured in a Lake Shore
model 7229 Susceptometer/Magnetometer. Resistance as a function of
temperature and magnetic field was measured using a standard
four-point probe technique in magnetic fields up to 5 T.

\section{Results and discussion}

\subsection{X-ray diffraction study}
\label{xray} Bulk La$_{1-x}$Ca$_{x}$MnO$_3$ has a distorted
perovskite structure, which is believed to be orthorhombic below
about 750 K \cite{kim,dai,hibble,das,aken}. The deformation of the
cubic perovskite lattice is determined by rotation of MnO$_6$
octahedra and the Jan-Teller distortion \cite{kim,aken}. For the
concentration range $0.25 < x < 0.5$, Jan-Teller distortion is
found to be negligible \cite{hibble,aken}. In the orthorhombic
space group {\it Pnma}, lattice constants are $a \approx
\sqrt{2}\,a_{p}$, $b \approx 2a_{p}$, and $c\approx
\sqrt{2}\,a_p$, where $a_p$ is the lattice constant of the
pseudo-cubic perovskite lattice. The symmetry of distortions is,
however, still a matter of some controversy \cite{lebedev}. In
some studies a monoclinic lattice is found \cite{lebedev,wollan}.
Generally, however, the deviations from cubic symmetry are found
to be fairly small, especially in the CMR range ($0.2 < x < 0.5$).
Several scientific groups have found only cubic symmetry in this
compound with lattice parameter $a_p$ in the range 0.385-0.388 nm
for $0.2 < x < 0.5$ (see, for example, Refs. \cite{mahen,laiho}).
\par
In this study, XRD investigation was used (along with other
methods) to judge the crystal perfection and structural
homogeneity of the samples. The x-ray spectra from powders and
bulk sections of different parts of the crystal were measured. We
found that XRD powder patterns from the outer and inner parts of
the crystal are quite different. The positions of all intense and
well-defined lines for the outer part of the crystal are exactly
consistent (within the uncertainty of the XRD recording) with
those of perfect cubic perovskite\footnote{The typical linewidth
(on the half of the magnitude) of the most intensive lines is
found to be in the range of 0.3--0.4 degree. The step of the XRD
pattern recording was 0.03 degree, that permitted to record surely
the line profiles and reveal available line splitting, which was
really found in the inner part of the crystal, as it is shown
below.}. A lattice constant $a_p = 0.38713$~nm was computed with
the corresponding unit cell volume $v_p \approx 0.0580$~nm$^3$. An
application of the Powdercell program \cite{pc}, which takes into
account not only positions, but intensities and profiles of the
lines as well, has shown that an orthorhombic {\it Pnma} lattice
also corresponds fairly well to the experimental XRD powder
pattern. This gives the lattice constants $a = 0.54849$~nm,
$b=0.77587$~nm, and $c=0.54645$~nm with a unit cell volume $v_o
\approx 0.2322$ nm$^3$.
\par
A quite different XRD powder pattern was obtained for inner part
of the crystal (Fig.~1). In low resolution this pattern looks the
same as that for the outer part of the crystal, but closer
inspection shows that most of the lines are split (into two,
three, or even four closely located lines). Significant line
splitting with a severely distorted cubic perovskite lattice was
found previously in low doped LaMnO$_3$ with Mn$^{4+}$ (or Ca)
concentration in the range 0-25~\% \cite{wollan}, which was
attributed to monoclinic distortions. Splitting due to small
orthorhombic distortions for the Ca concentration range
$0.25<x<0.5$ is usually very narrow, so that it cannot be
observed clearly (only some line broadening can be seen)
\cite{dai,li,he}.
\par
The splitting of the Bragg reflections could be also ascribed to
twinning, which is quite common in mixed-valence manganites.
Origin of twinning in perovskite crystals is well known
\cite{aken}. At high temperatures (above about 900$^{\circ}$C)
La$_{1-x}$Ca$_{x}$MnO$_3$ ($0.2 <x<0.5$) has a cubic perovskite
lattice \cite{he}. When cooling, the cubic-orthorhombic (or even
cubic-monoclinic) transition takes place, during which MnO$_6$
octahedra are rotated and distorted \cite{aken}. Since different
directions of rotation with respect to the original cubic axes are
possible, most La$_{1-x}$Ca$_{x}$MnO$_3$ single crystals are
twinned. In this case additional reflections (line splitting) can
appear if lattices in twin domains do not coincide. This can take
place in the case of monoclinic distortions \cite{aken} or owing
to other reasons (for example, when inhomogeneous strains are
present). It is impossible, however, on the basis of powder
diffraction data to see splitting due to twinning \cite{aken}.
Only special methods of single-crystal x-ray diffraction have
opportunities to reveal twinning effects in manganites
\cite{aken,tamaz}.  But the line splitting in a powder XRD pattern
due to considerable distortions from the cubic symmetry is quite
possible.
\par
We have applied the  above-mentioned Powdercell program to the
XRD data for inner part of the crystal. Assuming that crystal
lattice is orthorhombic, we obtained lattice constants $a =
0.54972$~nm, $b=0.77801$~nm, and $c=0.54551$~nm with a unit cell
volume $v_o \approx 0.2332$ nm$^3$, but agreement between the
calculated and experimental XRD patterns is far worse in this
case than that for the outer part of the crystal. It appears
therefore that distortions from cubic symmetry are more severe
than orthorhombic ones and they can be monoclinic as it was found
in some manganites in Refs. \cite{lebedev,wollan,tamaz}. This
suggestion has found a partial support after application of a
Monte Carlo program McMaille V3.04 \cite{mcmaille} to the
experimental XRD data. This has shown rather convincingly that
monoclinic lattice is much better corresponds to the XRD data of
the inner part of the sample than orthorhombic lattice.
\par
X-ray study with a Laue camera revealed that the XRD pattern of  a
small piece taken from the central part of the sample corresponds
to a single crystal structure. Investigation with an optical
microscope has shown, however, that the sample  as a whole is not
a single crystal but consists of a few grains.

\subsection{Magnetic properties}
Distinctions between the XRD patterns for outer and inner parts of
the crystal studied imply that their magnetic properties can be
different as well. That is indeed found in this study. The
temperature behavior of the ac susceptibility is presented in Fig.
2. It is seen that the magnetic transition from the paramagnetic
(PM) to the ferromagnetic (FM) state with decreasing temperature
is considerably smeared for the outer part of the crystal
indicating that this part is rather disordered. In contrast to
this, the magnetic transition in the central part of the sample is
fairly sharp, indicating a significant crystal perfection and
stoichiometric homogeneity for this part. The same conclusion can
be arrived from comparison of the temperature dependences of the
dc magnetization, $M(T)$, from which only that for central part of
the sample is shown in Fig. 3.
\par
Taking $T_c$ as the temperature of the inflection point in the
$M(T)$ curve, we found that $T_c \approx 216$~K.  Nearly the same
value can be obtained if $T_c$ were defined as the temperature at
which $M$ comes to half of the maximum value. This $T_c$ value
(216 K) is less than that ($T_c \approx 250$~K) usually found in
polycrystalline ceramic samples in La$_{1-x}$Ca$_{x}$MnO$_3$ with
$0.3 <x<0.35$, and which is indicated in the accepted phase
diagram for this system \cite{kim}. This value agrees, however,
well with those found in single crystal samples of the same
composition, where $T_c$ values are usually found to be in the
range 216--230 K \cite{chun,lyanda,hong,shin,tian}, although some
rare exceptions are known as well \cite{adams}.
\par
The inset in Fig. 3 shows the magnetic field dependence of the
magnetization at $T= 10$~K. It is seen that the magnetization is
close to saturation above the field $H\approx 0.3$~T. We have
recorded $M(H)$ dependences for other temperatures as well in the
range 4--51 K and at fields up to 9 T. Taking the saturation value
of $M$ (95.3 emu/g) at the highest field (9 T) in low temperature
range, we have obtained the magnetic moment per formula unit in
the sample studied to be equal to $\mu_{fu} \approx
3.65$~$\mu_{B}$, where $\mu_B$ is the Bohr magneton. For the
nominal composition La$_{1-x}$Ca$_{x}$MnO$_3$ ($x=0.33$) of the
sample studied, taking into account that the spin of Mn$^{+3}$ is
$S=2$ and that of the Mn$^{+4}$ is $S=3/2$, $\mu_{fu}$ should be
equal to $(4-x)$~$\mu_B$, that is to 3.67 $\mu_B$. The value of
$\mu_{fu} \approx 3.65$ $\mu_B$ obtained provides evidence that
the inner part of the sample is close to the nominal composition
and does not have any appreciable oxygen deficiency. Since the
inner part of the sample appears to be more homogeneous and
perfect than the outer part, resistive and magnetoresistive
measurements were done for inner part of the sample only.

\subsection{Resistive and magnetoresistive properties}
\label{resist}
 Temperature dependences of the resistivity, $\rho$,
and MR, $\delta_{0} = [R(H)-R(0)]/R(0)$, at different applied
fields, are shown in Figs.~4 and 5. Consider first the $\rho (T)$
behavior (Fig.~4) which appears somewhat complicated.  In doing so
it is appropriate to take a look at the behavior of the
temperature coefficient of resistivity (TCR) [defined as
$(1/R)(dR/dT)$] as well. This is presented in Fig.~6. Although the
$\rho (T)$ behavior reflects, generally, a rather high perfection
of the crystal studied, some quite distinct features in it should
be attributed to the influence of structural and magnetic
inhomogeneities. The $\rho (T)$ curve has two peaks, a sharp one
at $T_{p} \approx 225$~K, which is near the Curie temperature $T_c
\approx 216$~K, obtained from magnetization and susceptibility
studies (Figs. 2 and 3) and a far less pronounced one at
$T_{pin}\approx 190$~K. Additionally, $\rho (T)$ exhibits two
shallow minima, one is situated at $T_{min}^{'}\approx 201$~K
between the two peaks, and another in the low-temperature range at
$T_{min}^{''}\approx 16$~K (see inset in Fig.~4).
\par
Above $T_c$, the sample is a PM non-metal with an activated
temperature dependence of resistivity, following $\rho (T) \propto
\exp(E_a/T)$. The activation energy $E_a$ is found to be equal to
0.09 eV in agreement with that (about 0.1 eV) reported by other
authors for single crystal and ceramic samples of
La$_{1-x}$Ca$_{x}$MnO$_3$ ($x\approx 0.33$)
\cite{mahen,chun,lyanda,vert}. It is known
\cite{ramirez,coey,dagotto} that the conductivity of Ca-manganites
increases enormously at the transition to the FM state. In the
La$_{1-x}$Ca$_{x}$MnO$_3$ ($0.2\leq x\leq 0.5$), the PM-FM
transition occurs simultaneously with the insulator-metal one. For
this reason, the $\rho (T)$ dependence has a peak at a temperature
$T_p$.  In samples with fairly perfect crystalline structure,
$T_p$ is close to $T_c$. This is true to a considerable degree for
the sample studied. Other important measures of crystal perfection
in manganites are the ratio of the peak resistivity, $\rho (T_p)$,
and the residual resistivity at low temperature, $\rho (0)$. The
value of $\rho(T_p)/\rho(0)$ is about 200 for the sample studied.
This is a high value, comparable with that of single crystals and
epitaxial films of the best quality
\cite{coey,chun,lyanda,vert,martin,goyal}. For ceramic samples,
prepared by the solid-state reaction method, the ratio
$\rho(T_p)/\rho(0)$ is usually far less (see Refs.
\cite{kim,mahen,laiho}. It is well known that samples prepared by
the floating-zone method generally have a much sharper resistive
transition on going from the PM to FM state than is observed for
manganite films (compare data in Refs.
\cite{coey,chun,lyanda,vert,martin,goyal}). This is so indeed in
the sample studied which has an extremely sharp resistive
transition near $T_c$ (Fig. 4). This can be further illustrated by
the temperature behavior of TCR (Fig. 6). The highest value of TCR
near $T_c$ found in this study is about 38 \%/K; whereas, no more
than about 10 \%/K is found in the best quality
La$_{1-x}$Ca$_{x}$MnO$_3$ ($x\approx 0.3$) films
\cite{goyal,boris}.
\par
Along with general characteristics indicating a rather good
crystal perfection, the $\rho(T)$ behavior reveals, at the same
time, some features which are determined by sample inhomogeneities
(structural and magnetic). These are the second resistance peak at
$T_{pin}\approx 190$~K and the shallow resistance minimum in the
low-temperature range. Consider possible reasons for this
behavior. It is believed presently that the PM-FM transition in
La$_{1-x}$Ca$_{x}$MnO$_3$  is of first order at $0.25 < x < 0.4 $
(see Ref. \cite{boris2} and references therein). It is found as
well that FM clusters are present well above $T_c$ while some PM
insulating clusters can persist down to a range far below $T_c$.
This implies that the PM-FM transition has a percolative
character. With decreasing temperature, the PM volume fraction
decreases and that of the FM fraction increases. Since the PM
phase is insulating and the FM one is metallic, some kind of
insulator-metal transition takes place near $T_c$. The temperature
$T_p$ corresponds to the situation when metallic FM clusters have
merged together in a sufficient degree to ensure a decrease in
resistance with further temperature decrease. This temperature
indicates a transition from the insulating to the metallic state
and, therefore, is called conventionally the temperature of the
insulator-metal transition. Below $T_p$, the volume of the PM
phase continues to decrease with temperature causing a further
resistivity decreasing. The temperature width of this type of
transition depends in crucial way on the sample extrinsic
inhomogeneities induced by various technological factors during
the sample preparation (see Sec. \ref{int} and Ref.
\cite{boris2}). In samples, prepared by the floating-zone method,
these inhomogeneities are determined by mosaic blocks, twins,
inhomogeneous strains, and stoichimetric disorder. These defects
are present even in single-crystal Ca-manganites
\cite{muk1,aken,yuzhel}.
\par
Double-peaked (or shouldered) $\rho(T)$ curves similar to that
shown in Fig.~4  were often seen in polycrystalline manganite
samples \cite{vert,rao,wang,hernan,shyam}. In all cases this
behavior was quite reasonably attributed to the influence of grain
boundaries. The evident idea is that regions near the grain
boundaries are disordered and even depleted in charge carriers
compared with that inside the grains \cite{hernan,shyam}. This was
supported by direct experiments \cite{kar,bibes}, which have shown
that grain boundary regions (with thickness of the order 10 nm)
have a smaller $T_c$ than that in the core of the grain. For this
reason, when the cores of the grains become FM below $T_c$, the
grain boundary regions still remain in the paramagnetic,
insulating state, presenting barriers to the transport current.
This prevents the formation of an infinite FM cluster and causes
an increase in resistance with decreasing temperature when the FM
transition in cores of the grains is completed. The further
decrease in temperature, however, leads to a reduction of magnetic
disorder in the grain-boundary regions, eliminating some
insulating barriers. If this process proceeds sufficiently,
$\rho(T)$ behavior becomes metallic again with the result that a
second peak (or shoulder) appears in $\rho(T)$. This peak
indicates the appearance of the infinite FM cluster.
\par
The second peak at $T=T_{pin}$ in $\rho(T)$ curve (Fig.~4)
suggests the presence of some grain-boundary-like inhomogeneities
in the crystal studied. As a result the $\rho(T)$ dependence shown
in Fig. 4 reflects actually two resistive transitions in the
sample studied. The first is connected with PM-FM transition
inside the grains, while the second one signifies the magnetic
ordering of grain-boundary regions. On that ground the second peak
position at $T=T_{pin}\approx 190$~K can be taken as a rough
estimate of the Curie temperature $T_{cg}$ of the intergrain
regions.
\par
It should be noted that for weak intergrain connectivity (as often
takes place in ceramic samples) the second resistance peak can be
rather large and even higher than the first peak (caused by the
intragrain FM transition) \cite{rao,shyam}. Since the second peak
is rather weak for the sample studied, it can be said that
grain-boundary-like inhomogeneities are not as strong in it as in
ceramic samples. What can be the sources of such inhomogeneities
in samples prepared by the floating-zone method? First of all,
some rare grain boundaries cannot be excluded in this type of
sample \cite{vert}. But the major reason is that these samples
contain inevitably mosaic blocks and twin domains
\cite{aken,yuzhel}. Twin boundaries act effectively in the same
way as grain boundaries, presenting tunnel barriers \cite{yuzhel}.
The twins are present even in polycrystalline manganites prepared
by the solid-state reaction method, where large enough grains (a
few microns in size) contain multiple twins \cite{chen}.
\par
If the second peak in $\rho(T)$ is determined by the
grain-boundary-like interlayers with Curie temperature different
from that inside the grains, it must be depressed by an applied
magnetic field. This really occurs (see Fig.~4). At the same time
no visible features associated with grain boundaries can be seen
in the $M(T)$ dependence (Fig.~3). This is reasonable if the
volume fraction of the grain boundaries is sufficiently small.
\par
When the FM transition in grain/twin boundary regions is
completed, the boundaries can still exert an influence on the
sample conductivity. These boundary regions are places of not only
magnetic, but also structural disorder, where charge carriers can
be scattered. What's more, grain boundaries appear as natural
places for FM domain boundaries. On the basis of the known
experimental data \cite{kar,bibes} the following model of grain
boundaries in the manganites can be assumed.  A transition region
between any two grains consists of two layers depleted of charge
carriers (and having a reduced Curie temperature $T_{cg}$) and
some thin (thickness about a nanometer) insulating layer between
them. The latter presents the tunnel barriers for the charge
carriers even at low temperature $T < T_{cg}$ when the two
depleted layers become ferromagnetic and, most likely, metallic.
It is quite reasonable to assume that the thickness of the
insulating layers (tunnel barriers) is not the same throughout the
sample, so the system is percolating in this context as well.
\par
Quantum tunneling of the charge carriers occurs between states of
equal energy. Actually, however, there is always some energy level
mismatch between states in neighboring grains for different
reasons. In this case, a charge carrier should gain some energy
(for example, from phonons) to accomplish tunneling. The
intergrain conductivity is conditioned, therefore, by the two
processes: the tunneling and thermal activation. If the activation
energy of tunneling, $E_t$, is rather low, then tunneling at high
enough temperature, $kT>E_t$, is non-activated. In that event the
system behaves like a metal ($dR/dT>0$). Since the grain boundary
thickness is not the same throughout the system, the conductivity
is percolating. It is determined by the presence of ``optimal''
chains of grains with maximum probability of tunneling for
adjacent pairs of grains forming the chain. These ``optimal''
chains have some weak links (high-resistance tunnel junctions). At
low enough temperature the relation $kT<E_t$ can become true for
such links, and, hence, the measured conductivity of the system
will become activated. This is an evident reason for the observed
resistance minimum at $T_{min}^{''}\approx 16$~K (Fig.~4). More
generally it can be said that the minimum is determined by
competition between the intragrain conductivity and the intergrain
tunneling. The low-temperature resistance minimum is typical for
systems of FM regions with rather weak interconnections. It has
been observed in polycrystalline \cite{andres} and even single
crystal \cite{yuzhel} samples. We found that an applied magnetic
field has no noticeable effect on the resistance minimum position.
This may signify that the tunnel barriers are non-magnetic in the
low temperature range.
\par
It follows from discussion above that the double-peaked feature
and the low-temperature resistance minimum in $\rho(T)$ of the
sample studied can be adequately explained by the influence of the
grain/twin boundaries. The measured temperature dependences of MR
(Fig.~5) reflect the influence of these structural inhomogeneities
as well. Generally, the MR in manganites is determined by
intrinsic and extrinsic causes. The CMR is an intrinsic effect
determined by the ability of an external magnetic field to
increase the magnetization. It is clear that at low temperature
($T\ll T_c$) when all spins are already aligned by the exchange
interaction, this ability is minimal. The possibility to
strengthen the magnetic order with an external magnetic field
increases profoundly near $T_c$ where the magnetic order becomes
weaker. For this reason, CMR is maximal near $T_c$ and goes to
nearly zero value for $T \ll T_c$. This behavior is unique to bulk
or film manganites of rather good crystal perfection
\cite{vert,boris}. The presence of extrinsic inhomogeneities (such
as grain boundaries) gives rise to an extrinsic MR effect.
Considering Fig.~5, it is safe to assume that the sharp MR peak at
$T\approx 220$~K is caused by CMR; whereas, the rather weak peak
at $T\approx 173$~K is connected with the influence of grain/twin
boundaries. The measured temperature behaviors of the MR (Fig.~5)
and the resistance (Fig.~4) are clearly correlated and both of
them reflect the influence of extrinsic inhomogeneities. At the
same time, they also demonstrate the good crystal perfection of
the sample. Really, the maximum value of MR, expressed as
$\delta_{0} = [R(H)-R(0)]/R(0)$, is about -96.4 \% near the Curie
temperature in a 5 T field. This is the CMR indeed. Since
$\delta_{0}$, by definition, cannot be higher than 100 \%, another
characterization of CMR, namely, $\delta_H = [R(0)-R(H)]/R(H)$ is
frequently used. In the crystal studied, the maximum $\delta_H$ in
the 5 T field is about $\approx 2680$~\%. This is even higher than
the MR found in some single-crystal Ca-manganites of the nearly
same chemical composition (see Ref. \cite{lyanda}). In ceramic
samples, however, the values of $\delta_H$ are usually about two
orders of the magnitude less \cite{mahen,rao,kar}.
\par
It is known that a contribution to MR coming from
grain-boundary-like inhomogeneities increases with decreasing
temperature. Discussion of the possible mechanisms for this
extrinsic type of the MR can be found in Refs.
\cite{gupta,hwang,evetts,ziese}. The sample studied does indeed
show a continuous increase in MR with decreasing temperature in a
0.5 T field in the temperature range below 200 K (Fig.~5). In
higher fields, the MR behavior is more complicated. It should be
noted that the temperature behavior of MR and its changes with
increasing applied magnetic field, revealed in the sample studied
(Fig.~5), are quite similar to that found in Ref. \cite{vert} for
the MR of \emph{a single grain boundary} in a Ca-manganite crystal
grown by the floating-zone method. This suggests once again that
grain/twin boundaries in the sample studied are not large in
number.
\par
It is well established \cite{gupta,hwang,evetts,ziese,hwang1} also
that the extrinsic MR connected with grain boundaries has one more
characteristic feature. In the low-field region ($H < H_s$, where
$H_s$ is a characteristic field) the resistance decreases rapidly
with increasing $H$  [so called, low-field MR (LFMR)]. For $H>
H_s$, the resistance changes more gradually [high-field MR
(HFMR)]. In both these field regions, the resistance changes
almost linearly with $H$. The LFMR is found to decrease rather
rapidly with increasing temperature  so that the broken-line type
of the $R(H)$ dependence is changed to a smooth featureless
dependence for higher temperature. All these features of extrinsic
MR can be seen in Fig. 7, where MR {\it vs} $H$ for two
temperatures (4.2 K and 68 K) is presented. In the known studies
of manganites the characteristic field $H_s$ is typically below
0.5~T  and is attributed to the field of magnetic domain rotation
(when the magnetization magnitude comes close to the technical
saturation value). This corresponds roughly to the dependence of
$M(H)$ shown in the inset to Fig. 3.
\par
It is currently believed \cite{hwang,ziese,hwang1} that the
grain-boundary MR in manganites is determined primarily by the
spin-dependent tunneling of charge carriers between the grains.
The LFMR is attributed to magnetic alignment of the grains;
whereas, HFMR to an increase in the spontaneous magnetization in
higher field. For the low temperature range the latter process
occurs mainly within the disordered grain-boundary regions, while
for high enough temperature it takes place mainly inside the
grains causing CMR. This competition of CMR and the extrinsic MR
gives a somewhat intricate picture of the total MR behavior found
in this study. We will not consider further these or other
peculiarities of the extrinsic MR connected with
grain-boundary-like inhomogeneities nor the corresponding possible
mechanisms of this behavior. It is clear enough from the data
obtained that the sample studied contains inhomogeneities of this
type.
\par
In conclusion, we have studied structure, magnetic, and
magnetoresistive properties of crystals with nominal composition
La$_{0.67}$Ca$_{0.33}$MnO$_3$ prepared by the floating zone
method. The properties indicate undoubtedly that the sample is of
fairly high crystal perfection. In particular, it has very low
resistivity at low temperature and a huge MR near the Curie
temperature ($\delta_H = [R(0)-R(H)]/R(H)$  in the field $H=$~5 T
is about $\approx 2680$~\%). At the same time, some features of
the transport and magnetoresistive properties reflect the
essential influence of extrinsic grain-boundary type structural
inhomogeneities arising due to technological factors in the sample
preparation. It is shown that these inhomogeneities are rare
grains and twins.
\par The work at TAMU was supported by the Robert
A. Welsh Foundation, Houston, Texas (Grant A-0514) and the
National Science Foundation (DMR-0111682 and DMR-0422949). B. I.
B. acknowledges support from Program ``Nanostructural systems,
nanomaterials, nanotechnologies'' of the National Academy of
Sciences of Ukraine under grant No. 3-026/2004.



\newpage

\centerline{\bf{Figure captions}} \vspace{12pt}

Figure 1. XRD powder pattern of the inner part of the crystal with
nominal composition La$_{0.67}$Ca$_{0.33}$MnO$_3$. The indicated
indexes correspond to lines for cubic perovskite-like unit cell.
It is seen that most of the lines are split. To show this more
clearly, the region around the (211) reflection is magnified in
the inset. \vspace{12pt}

Figure 2. Temperature dependences of the real part, $\chi^{'}$, of
ac susceptibility for the outer (a) and central (b) parts of the
crystal studied. The dependences were recorded (with a 125~Hz ac
magnetic field $H_{ac}=1\times 10^{-5}$~T) with increasing
temperature after the samples were cooled in zero field.
\vspace{12pt}

 Figure 3. Temperature dependences of the magnetization of the
central part of the sample recorded (in a dc magnetic field
$H_{dc}=0.2$~mT) with increasing temperature after the sample was
cooled in zero field. The inset shows the magnetic-field
dependence of the magnetization at $T=10$~K. \vspace{12pt}

Figure 4. Temperature dependences of the resistivity of the
sample, recorded in zero magnetic field and in fields $H=$~0.5, 1,
2, 3.5 and 5 T. Inset shows a shallow resistance minimum at low
temperature. The minimum temperature, $T_{min}^{''}$, is about 16
K.  \vspace{12pt}

Figure 5. Temperature dependences of the magnetoresistance of the
sample, recorded at different magnitudes of applied magnetic
field. \vspace{12pt}

Figure 6. Temperature behavior of temperature coefficient of
resistance, $(1/R)(dR/dT)$, of the sample at zero magnetic field.
\vspace{12pt}

Figure 7. Magnetic-field dependences of the magnetoresistance,
$[R(H)-R(0)]/R(0)$, of the sample at $T=4.2$~K and 68 K. The
dependences were recorded for increasing and subsequently for
decreasing applied magnetic field. No significant hysteresis in
the curves can be seen.

\newpage

\begin{figure}[htb]
\centering\includegraphics[width=0.85\linewidth]{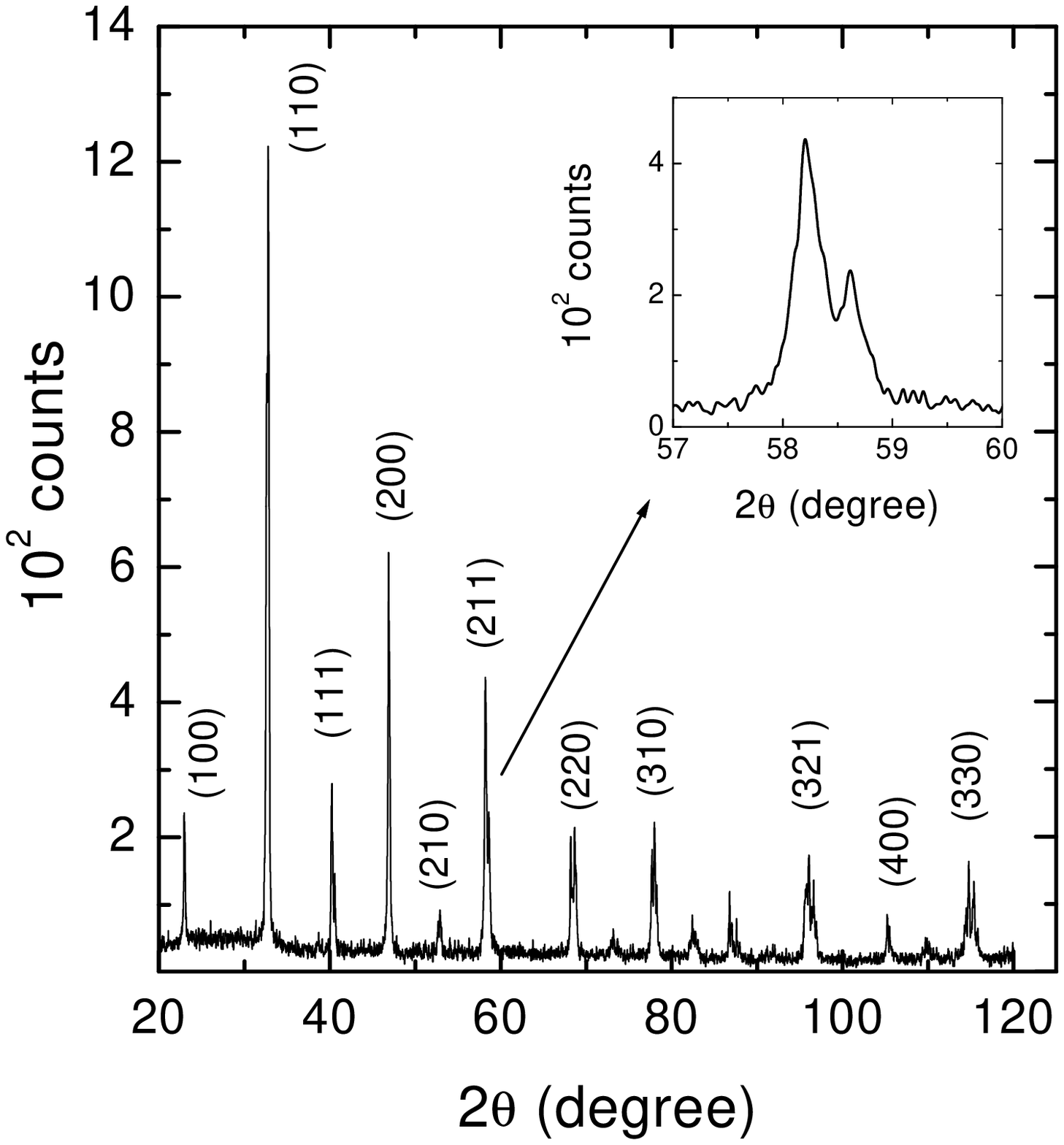}
\centerline{Figure 1 to paper Belevtsev et al.}
\end{figure}

\begin{figure}[htb]
\centering\includegraphics[width=0.85\linewidth]{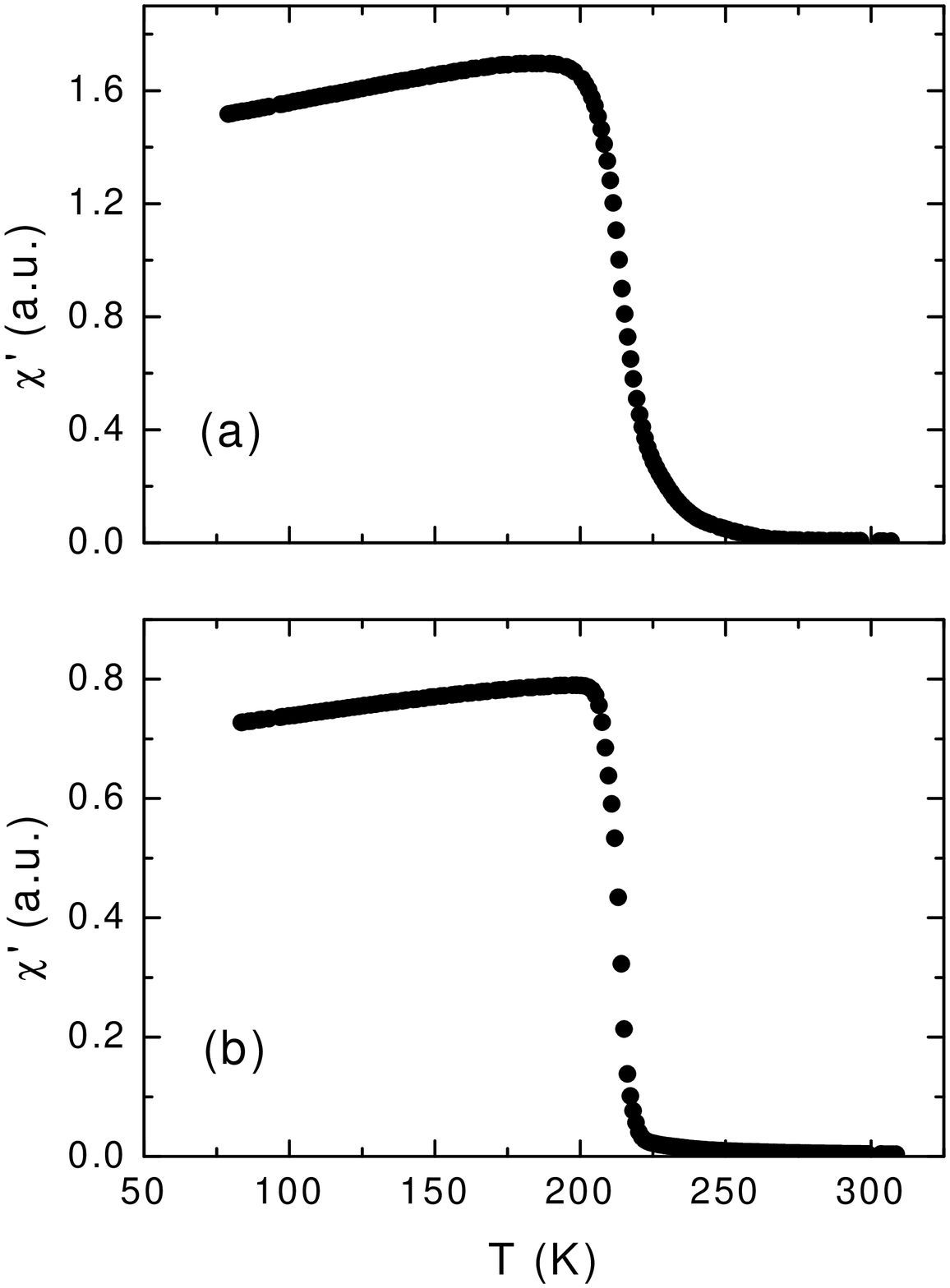}
\centerline{Figure 2 to paper Belevtsev et al.}
\end{figure}

\begin{figure}[htb]
\centering\includegraphics[width=0.85\linewidth]{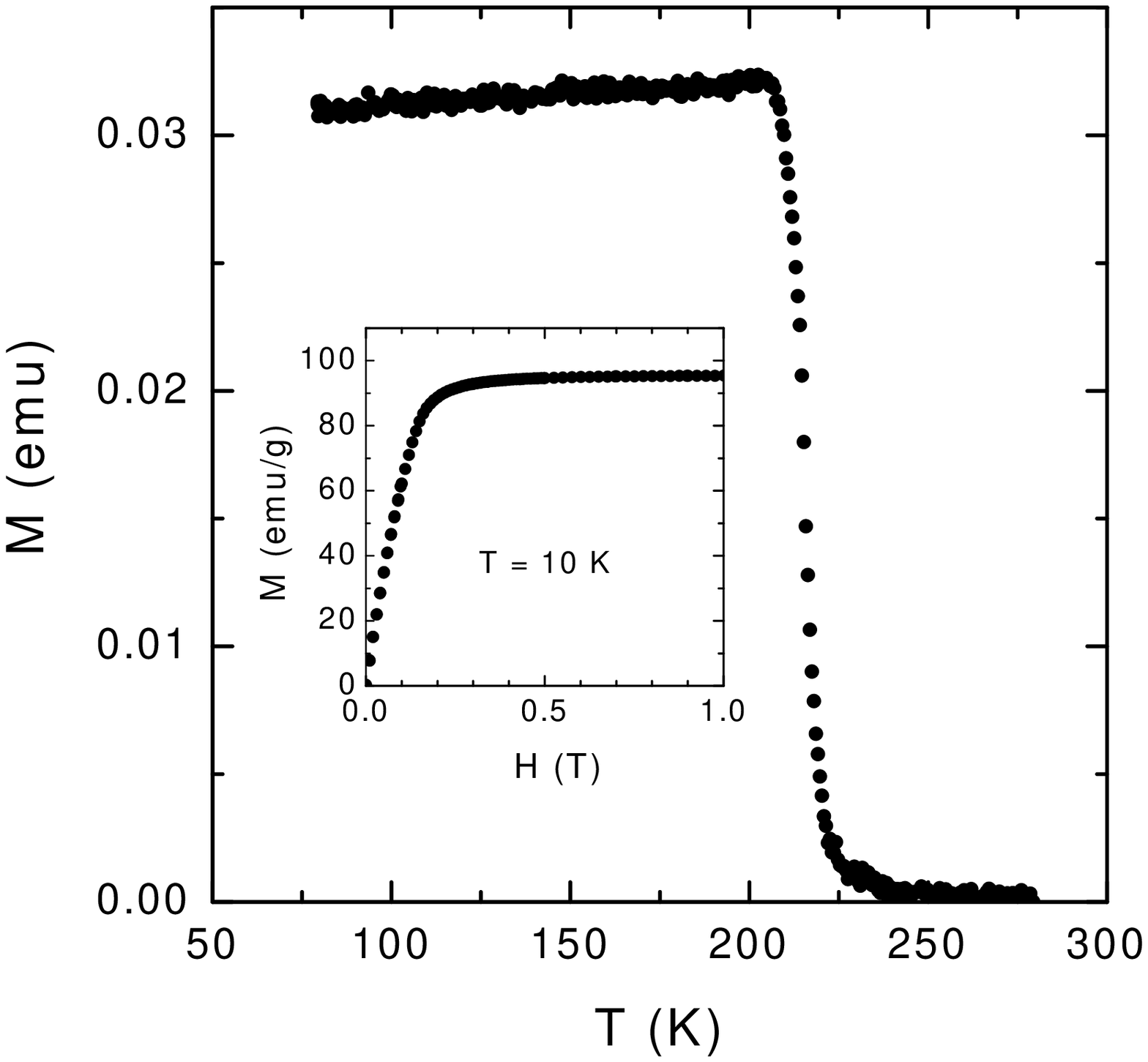}
\centerline{Figure 3 to paper Belevtsev et al.}
\end{figure}

\begin{figure}[htb]
\centering\includegraphics[width=0.85\linewidth]{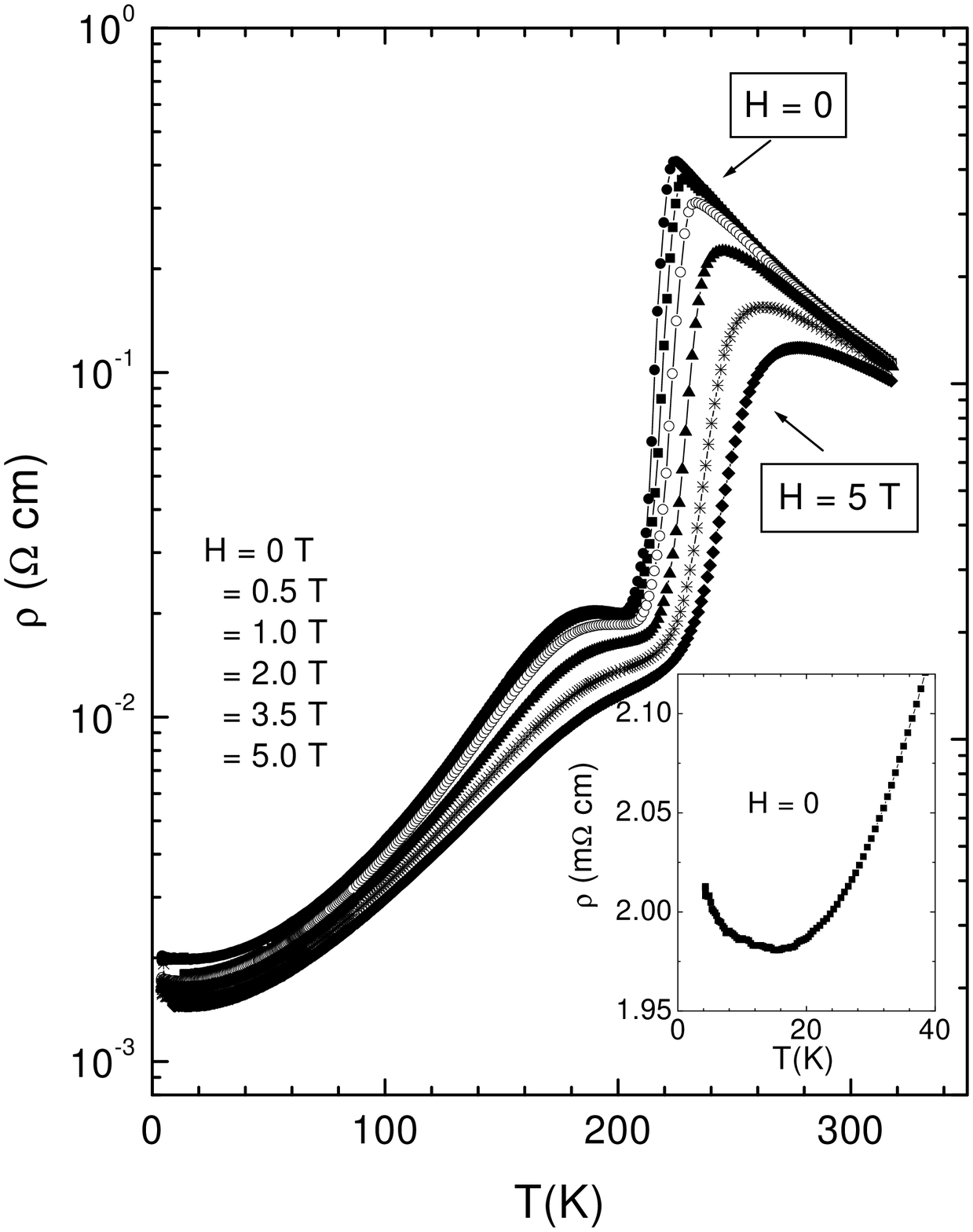}
\centerline{Figure 4 to paper Belevtsev et al.}
\end{figure}

\begin{figure}[htb]
\centering\includegraphics[width=0.85\linewidth]{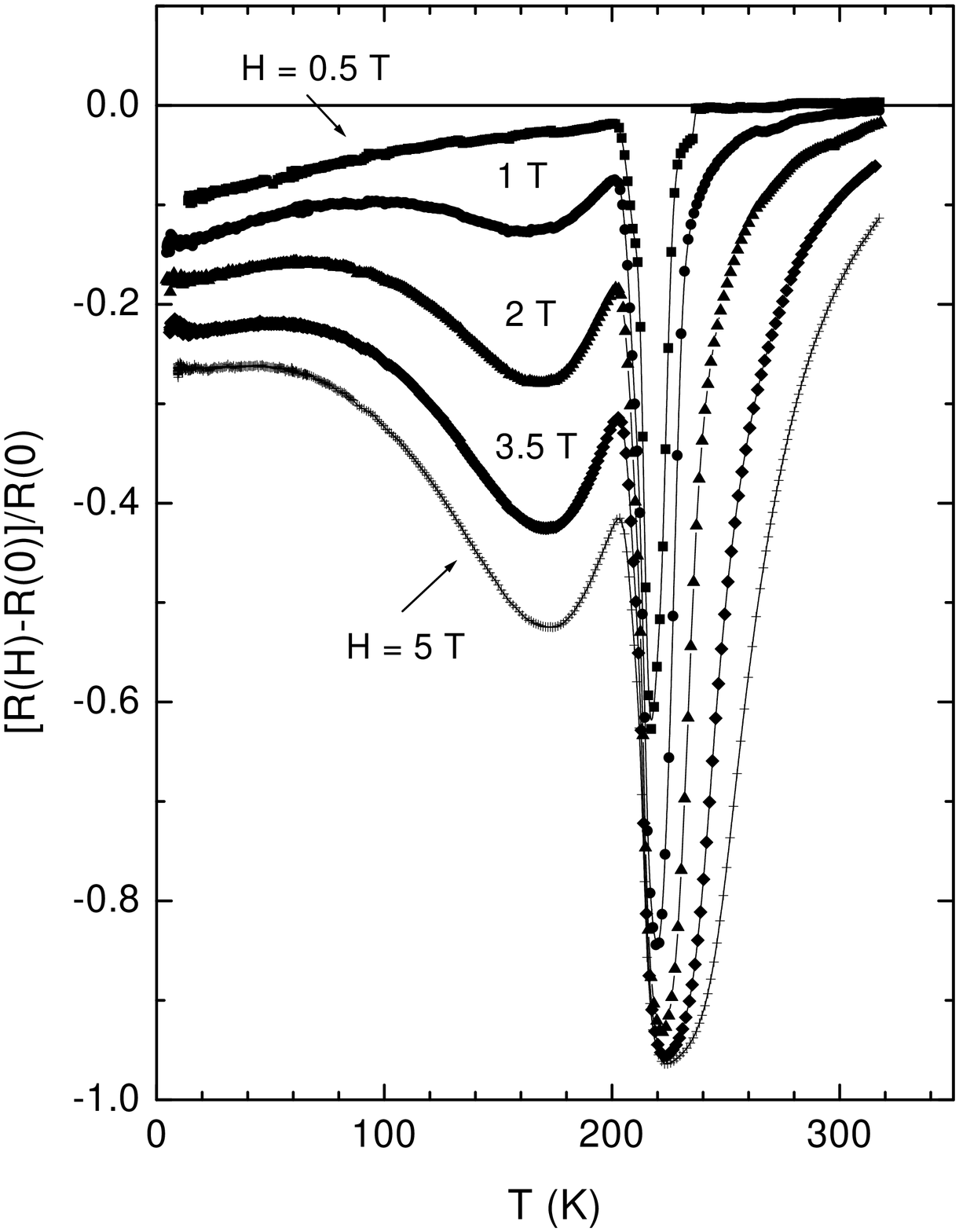}
\centerline{Figure 5 to paper Belevtsev et al.}
\end{figure}

\begin{figure}[htb]
\centering\includegraphics[width=0.85\linewidth]{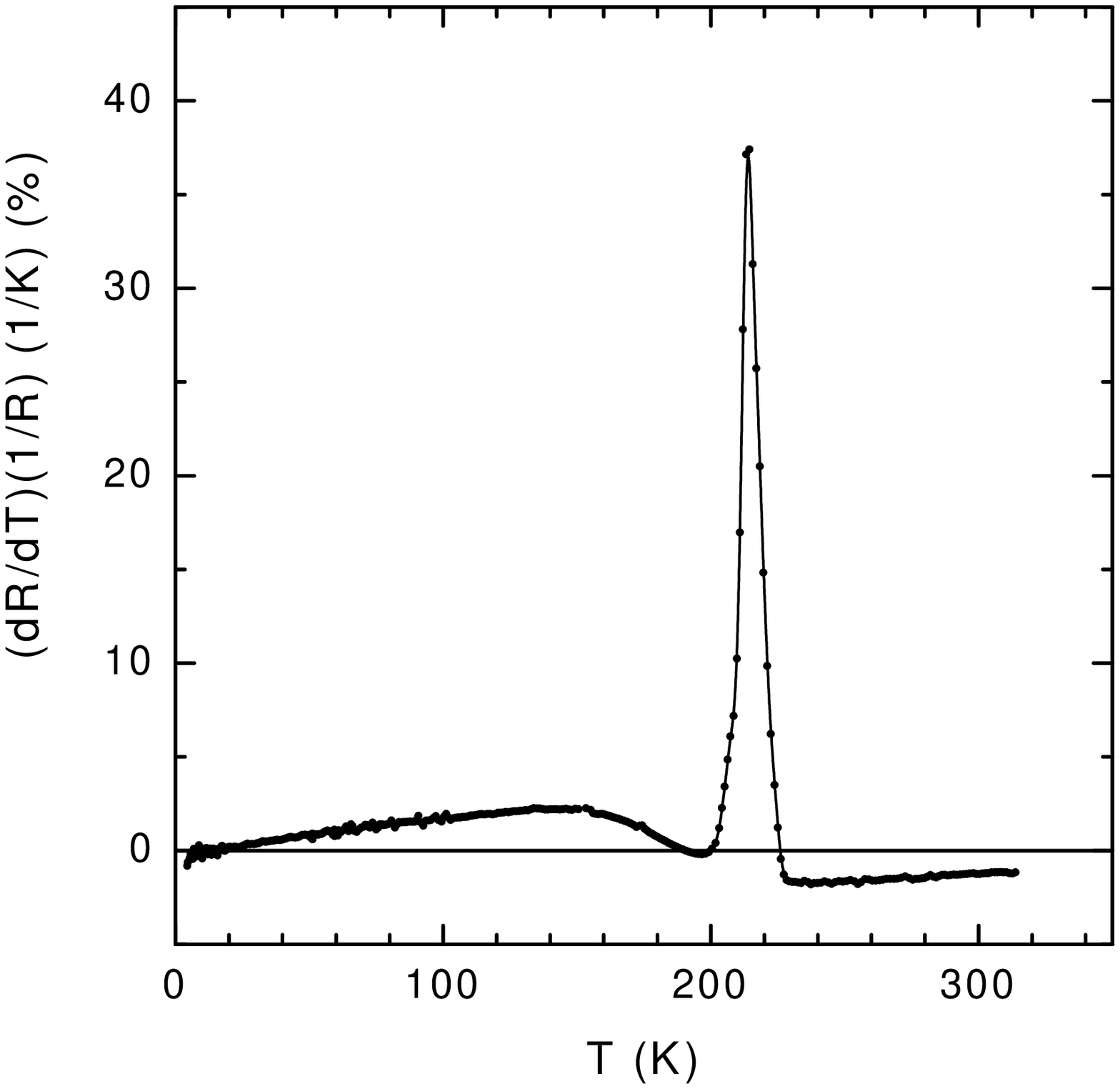}
\centerline{Figure 6 to paper Belevtsev et al.}
\end{figure}

\begin{figure}[htb]
\centering\includegraphics[width=0.85\linewidth]{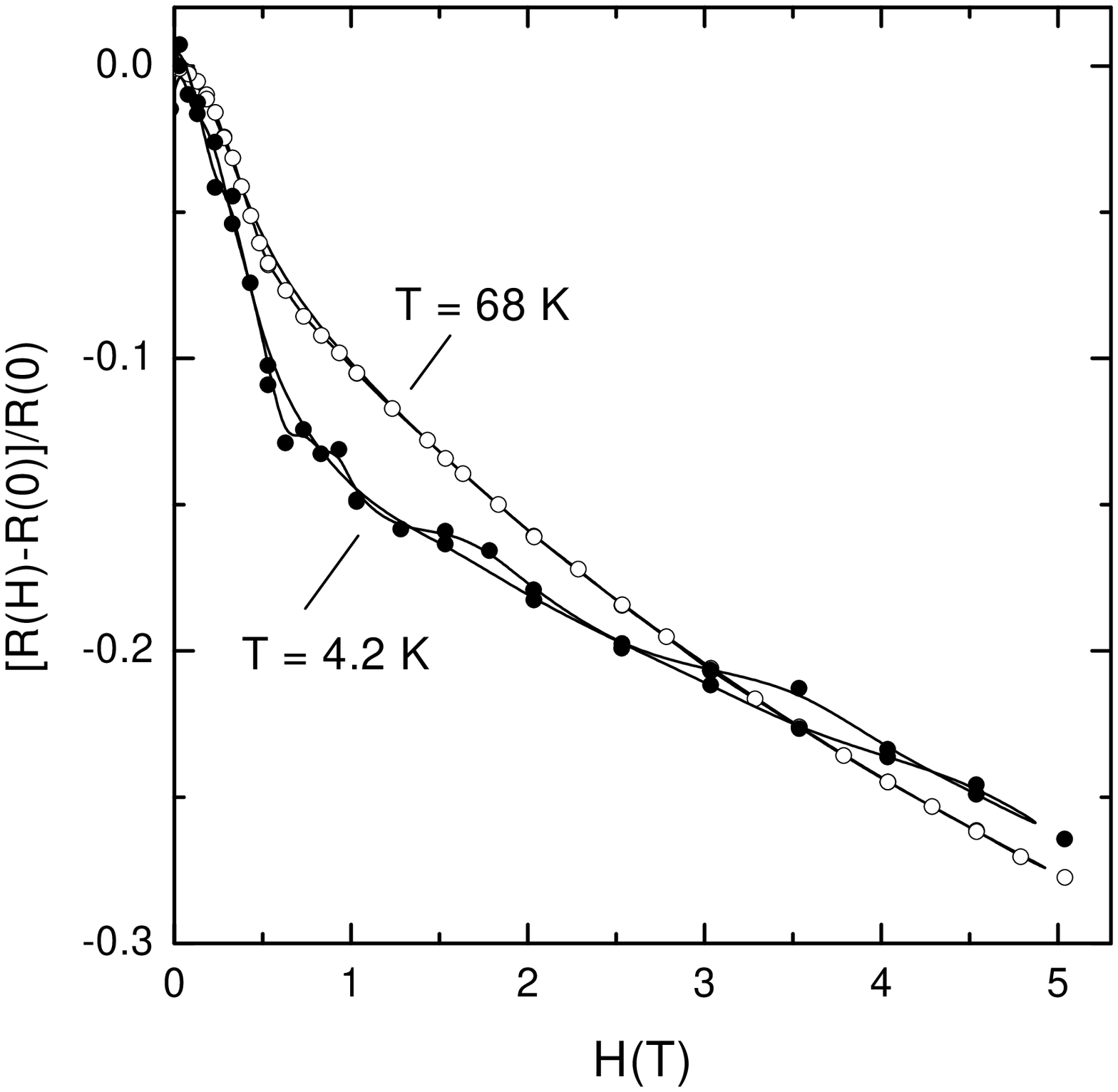}
\centerline{Figure 7 to paper Belevtsev et al.}
\end{figure}

\end{document}